\documentstyle[preprint,aps,epsfig]{revtex}
\renewcommand{\theequation}{\arabic{section}.\arabic{equation}}
\tighten
\begin{document}
\draft

\title{Deconstructing the vertex Ansatz\\
in three dimensional quantum electrodynamics}
\author{ C.\ J.\ Burden and P.\ C.\ Tjiang\vspace*{0.2\baselineskip}}
\address{Department of Theoretical Physics, 
Research School of Physical Sciences and Engineering,\\
 The Australian National University, Canberra ACT 0200 , AUSTRALIA 
\vspace*{0.2\baselineskip}\\}
\maketitle

\begin{abstract}
We consider the problem of designing an Ansatz for the fermion-photon 
vertex function, using three-dimensional quantum electrodynamics as a 
test case.  In many existing studies, restrictions have been placed on 
the form of the vertex Ansatz by making the unsubstantiated assumption 
that in the quenched, massless limit the Landau gauge Dyson-Schwinger 
equations admit a trivial solution.  We demonstrate, without recourse 
to this assumption, the existence of a non-local gauge in which the 
fermion propagator is the bare propagator.  This result is used to 
provide a viable Ansatz for part of the vertex function.   
\end{abstract}

\section{Introduction}

Dyson-Schwinger equations (DSEs) provide a viable method for studying the 
non-perturbative behaviour of field theories.  In gauge field theories in 
particular, a common technique is to analyse the fermion propagator by 
truncating the infinite tower of DSEs at the level of the propagator DSEs.  
One then implements an intelligent Ansatz for the gauge boson propagator 
and boson-fermion vertex function\cite{RW94}.  

For quantum electrodynamics in either three (QED3) or four (QED4) 
dimensions, there has for some years now 
been an ongoing programme of improving the Ansatz for the fermion-photon 
vertex\cite{CP91,BR93,DMR94,BP96,BKP98}.  A principal goal of this programme 
is to invent an Ansatz which respects the gauge covariance of Green's 
functions in accordance with the transformation properties discovered 
by Landau and Khalatnikov (LK)\cite{LK56}.  

It has traditionally been common practice in DSE studies of QED3 or QED4 
to assume, either implicitly or explicitly, that in the 
quenched (i.e. $N_f \rightarrow 0$), massless limit, the DSEs admit the 
trivial solution of bare fermion propagator and bare 
vertex in Landau gauge\cite{CP91,BR93,DMR94,BP96,K97,AMM97}.  In the 
case of QED4, this assumption has recently been questioned by by 
Bashir et al.\cite{BKP98}, who use the one-loop perturbative correction 
to the vertex\cite{KRP95} to model an unknown additional transverse part 
of the vertex.  

In this paper we explore the problem of constructing a viable vertex Ansatz 
for the case of QED3, without recourse to the above simplifying assumption.  
We choose to work with QED3 rather than 
its four dimensional counterpart because its benign ultraviolet 
properties render the integrals we encounter finite, 
obviating the need for awkward numerical regularisations.  The four 
component version of massless QED3 which we consider here has an 
interesting chiral-like $U(2)$ symmetry which, if dynamically broken, 
leads to dynamical mass generation.  The evidence from both DSE 
and lattice calculations, suggest that this is almost certainly the case, 
at least for small numbers of fermion flavours\cite{M96}.  Here, however, 
we shall be considering the chirally symmetric solutions, which must also 
respect the LK transformations, and can therefore be used to place 
restrictions on the allowed form of the vertex Ansatz.  

We shall demonstrate the existence of a gauge in which massless QED3 
admits a bare fermion propagator solution, though this may not be 
Landau gauge, even in the quenched case.  The vertex function 
decomposes into two parts, an in principle known part which reduces to 
the bare vertex in the gauge mentioned, and an extra, unknown transverse 
part.  By studying the gauge parameter dependence of the photon polarisation 
scalar, we construct a computationally viable vertex Ansatz for the first 
of these two parts.  

The massless fermion DSE for QED3 in Euclidean momentum space is
\begin{equation}
1=i\gamma \cdot p S(p)+e^2\int \frac{d^3q}{(2\pi)^3} D_{\mu \nu}(p - q) 
\gamma_\mu S(q) \Gamma_\nu (q,p) S(p),    \label{masslessDSE}
\end{equation}
where the Euclidean $\gamma$ matrices satisfy 
$\{\gamma_\mu,\gamma_\nu\}=2\delta_{\mu \nu}$. $D_{\mu \nu}$ is the photon 
propagator which, for the class of covariant, nonlocal gauges and for 
$N_f$ fermion flavours takes the form  
\begin{equation}
D_{\mu \nu}(k)  =   \frac{1}{1 + N_f\Pi(k^2)}D_{\mu \nu}^{\rm T}(k) 
              - k_\mu k_\nu \Delta(k^2),   \label{photon}
\end{equation}
where
\begin{equation}
D_{\mu \nu}^{\rm T}(k) =\frac{1}{k^2} 
   \left(\delta_{\mu \nu} - \frac{k_\mu k_\nu}{k^2}\right).   \label{DTdef}
\end{equation}
The gauge choice 
\begin{equation}
\Delta(k^2) = -\frac{\xi}{k^4},   \label{covgauge} 
\end{equation} 
with $\xi$ constant, defines the usual covariant gauge.  
The regulated photon polarisation scalar\cite{P92} is given by 
\begin{equation}
\Pi(p^2) = -\frac{e^2}{2p^2} 
\left( \delta_{\mu \nu} - 3 \frac{p_\mu p_\nu}{p^2} \right)
 \int \frac{d^3q}{(2 \pi)^3} {\rm tr} 
\left[ \gamma_\mu S(q + \mbox{$\frac{1}{2}$}p) 
\Gamma_\nu (q + \mbox{$\frac{1}{2}$}p,q - \mbox{$\frac{1}{2}$}p) 
      S(q - \mbox{$\frac{1}{2}$}p) \right].    \label{vacuum1}
\end{equation} 

A common starting point for constructing vertex Ans\"atze is the Ball and 
Chiu vertex\cite{BC80}.  Writing the dressed fermion propagator in the 
chirally symmetric phase as  
\begin{equation}
S(p) = \frac{1}{i \gamma \cdot p A(p^2)},  \label{fermion}
\end{equation}
the Ball-Chiu vertex is given by 
\begin{equation}
\Gamma_\mu (p,q) = \Gamma_\mu^{\rm BC} (p,q) + \Gamma_\mu^{\rm T} (p,q),  
                                       \label{general}
\end{equation}
where
\begin{equation}
\Gamma_\mu^{\rm BC} (p,q)  = 
     \frac{1}{2}\left[A(p^2)+A(q^2)\right] \gamma_\mu 
     + \frac{(p + q)_\mu}{p^2 - q^2} \left[A(p^2)-A(q^2)\right] 
    \frac{\gamma \cdot p + \gamma \cdot q}{2}  \label{BCvertex},
\end{equation}
and the transverse part of the vertex $\Gamma_\mu^{\rm T}$ satisfies
\begin{equation}
(p - q)_\mu \Gamma_\mu^{\rm T} (p,q) = 0, \hspace{5 mm} 
                       \Gamma_\mu^{\rm T}(p,p) = 0.  \label{tconstraint}
\end{equation}
Without loss of generality, the transverse part can be written in terms of 
eight scalar functions $g_i$ as 
\begin{equation}
\Gamma_\mu^{\rm T} (p,q) =  
\sum_{i=1}^8 T_\mu^i (p,q) g_i (p^2,q^2,p \cdot q) .  \label{transverse}
\end{equation}
Charge conjugation invariance restricts all $g_i$ to be symmetric under 
interchange of $p$ and $q$, except $g_6$, which is antisymmetric. 
Given the form of the fermion DSE, it is reasonable to assume that in 
the chirally symmetric sector, only those $T_\mu^i$ consisting of terms 
with odd numbers of Dirac matrices contribute to the vertex function.  
Any terms containing even numbers of Dirac matrices would need to 
conspire to produce contributions to the final term in Eq.~(\ref{masslessDSE})
which integrate to zero.  Throughout this paper we shall therefore assume 
that only the four $T_\mu^i$ listed in the appendix contribute to 
$\Gamma_\mu^{\rm T}$ in the chirally symmetric sector.  

We also adopt the position that a useful Ansatz should be such that 
the $g_i$ have no dependence on the gauge fixing function $\Delta(k^2)$ 
other than that entering implicitly via a functional dependence on other 
Green's functions (such as the $A$ dependence in $\Gamma^{\rm BC}$).  This 
allows one to exploit gauge covariance of the solutions of the DSE as a 
constraint on the functions $g_i$.  
\footnote{Suppose we allow the vertex Ansatz to have an explicit $\Delta$ 
dependence.  Then one can always impose gauge 
covariance on solutions of the DSEs by fiat as follows: First arbitrarily 
specify a vertex Ansatz $\Gamma_\mu(p,q;0)$ of the form of 
Eq.~(\ref{general}) to be the Landau gauge vertex, and then {\em define} 
the Ansatz $\Gamma_\mu(p,q;\Delta)$ in other gauges to be the LK transform 
of $\Gamma_\mu(p,q;0)$.  Form invariance of the DSEs under LK transformations 
then ensures that any propagator solution will automatically have the 
desired LK transformation properties.  Since our choice of Landau gauge 
vertex was arbitrary, a constraint demanding the gauge covariance 
of propagator solutions then becomes vacuous.}  

\setcounter{equation}{0}
\section{The bare propagator gauge}
 
The Ball-Chiu vertex is arrived at by requiring that the following 
criteria be satisfied:\\ 
(a) it should satisfy the Ward-Takahashi identity (WTI): 
$i(p - q)_\mu \Gamma_\mu(p,q) = S^{-1}(p) - S^{-1}(q)$, \\
(b) it should satisfy the Ward identity: 
$i\Gamma_\mu(p,p) = \partial_p S^{-1}(p)$, or equivalently, be free of 
kinematic singularities as $p^2 \rightarrow q^2$ and \\
(c) it must have the 
same transformation properties as the bare vertex $\gamma_\mu$ under 
Lorentz transformations and charge conjugation.  

In many papers dealing with QED DSEs 
\cite{CP91,BR93,DMR94,BP96,K97,AMM97} the vertex function is effectively  
assumed to be restricted by a further condition, namely:\\
(d) it must reduce to the bare vertex (i.e. $\Gamma_\mu^{\rm T}$ reduces 
to zero) when dressed propagators are replaced by bare propagators.

A consequence of condition (d) is that, in Landau gauge ($\Delta = 0$), 
the chirally symmetric solution to the massless, quenched fermion DSE 
Eq.~(\ref{masslessDSE}) is the free propagator 
\begin{equation}
S(p) = \frac{1}{i\gamma\cdot p}, \hspace{5 mm}\mbox{ or }\hspace{5 mm}
S(x) = \frac{\gamma\cdot x}{4\pi\left|x\right|^3}, \label{freeprop}
\end{equation}
where $S(x) = \int d^3p/(2\pi)^3\, \exp(-ip\cdot x) S(p)$ is the 
propagator in coordinate space and $\left|x\right| = (x_\mu x_\mu)^{1/2}$.  
Eq.~(\ref{freeprop}) follows from the identity 
\begin{equation}
\int \frac{d^3q}{(2 \pi)^3} D_{\mu \nu}^{\rm T} (p - q) 
    \frac{\gamma_\mu \gamma\cdot q \gamma_\nu}{q^2} = 0 
                                  \label{angid}
\end{equation}
which ensures that the perturbative one-loop fermion self energy in this 
case is zero.  

In reference \cite{BKP98} Pennington et al. argue that restriction (d)  
is not justified in QED4, and indeed not consistent with perturbation 
theory.  Likewise, we have no reason to assume in QED3 that condition 
(d) is valid or that the bare fermion propagator is a solution to the 
quenched theory in Landau gauge.  In fact, the vertex function is in 
principle determined by the vertex DSE 
\begin{equation}
\Gamma_{\mu ab}(p,q) = \gamma_{\mu ab} + \int \frac{d^3\ell}{(2\pi)^3} 
 \left[S(p + \ell)\Gamma_\mu(p + \ell,q + \ell)S(q + \ell)\right]_{dc}
       K_{cd,ba}(q + \ell,p + \ell,\ell),   \label{VDSE} 
\end{equation}
where the kernel $K$ is a sum of skeleton graphs containing the dressed 
fermion and photon propagators and dressed vertex function.  The photon 
propagator, and hence the kernel, is determined via Eq.~(\ref{vacuum1}) 
once the fermion propagator and vertex function are determined.  
Eq.~(\ref{VDSE}) is then an integral equation relating $\Gamma_\mu$ and 
$S$.  We therefore expect the full vertex to be functionally dependent 
on just the fermion propagator, and, more specifically, for both 
$\Gamma_\mu$ and $\Pi$ to be determined if the fermion propagator 
is specified to be the bare propagator.  

One is tempted to ask whether there may nonetheless be some (possibly 
nonlocal) gauge in which the bare propagator is the solution to the 
fermion DSE.  Such a scenario has been suggested by a number of authors 
in the context of the nonquenched theory\cite{K97,AMM97,S90}.  In these 
studies, however, the vertex Ansatz is taken either to be the bare vertex 
or a scalar function multiplied by the bare vertex.  In either case 
condition (d) is satisfied and in the quenched limit the desired gauge 
leading to a trivial solution is Landau gauge.  Here we demonstrate the 
existence of a nonlocal gauge in which the fermion DSE 
is solved by the bare propagator without assuming condition (d).  

Let us take the solution to the vertex DSE (\ref{VDSE}) to be  
\begin{equation}
\Gamma_\mu (p,q) = \gamma_\mu + \bar{\Gamma}_\mu^{\rm T}(p,q), \label{barver}
\end{equation}
when the fermion propagator is the bare propagator.  With 
$\bar{\Gamma}_\mu^{\rm T}$ transverse, this form is consistent with 
the Ball-Chiu form Eq.~(\ref{general}).  
Substituting Eqs.~(\ref{photon}), (\ref{freeprop}) and (\ref{barver}) 
into (\ref{masslessDSE}) we obtain 
\begin{eqnarray}
\lefteqn{\int\frac{d^3q}{(2\pi)^3} (p - q)_\mu (p - q)_\nu \Delta[(p - q)^2]  
     \frac{\gamma_\mu \gamma\cdot q \gamma_\nu}{q^2} } \nonumber \\
& = &  \int\frac{d^3q}{(2\pi)^3} 
     \frac{1}{\left\{1 + N_f\Pi\left[(p - q)^2\right]\right\}(p - q)^2} 
       \frac{\gamma_\mu \gamma\cdot q \bar{\Gamma}_\mu^{\rm T}(p,q)}{q^2} 
                                         \nonumber \\
& & +  \int\frac{d^3q}{(2\pi)^3}  \frac{1}{1 + N_f\Pi\left[(p - q)^2\right]}
 D_{\mu\nu}^{\rm T}(p - q) \frac{\gamma_\mu \gamma\cdot q \gamma_\nu}{q^2} . 
                                  \label{inteq}
\end{eqnarray}  
A solution of these equations for the scalar function $\Delta(k^2)$ will 
then provide us with a gauge in which the fermion DSE is solved 
by the bare propagator.  If $\bar{\Gamma}_\mu^{\rm T}$ 
includes just those $T_\mu^i$ whose terms contain an odd number of Dirac 
matrices, Lorentz covariance entails that the right hand side of 
Eq.~(\ref{inteq}) takes a form 
\begin{equation} 
- \gamma\cdot p \phi(p^2),  
\end{equation}
for some scalar function $\phi$.  Fourier transforming both sides then gives 
\begin{equation}
 \frac{\gamma_\mu \gamma\cdot x \gamma_\nu}{4\pi\left|x\right|^3}
   \partial_\mu \partial_\nu \Delta(x^2) = \gamma\cdot\partial\phi(x^2), 
\end{equation}
which simplifies to 
\begin{equation}
\frac{d}{du}\left[u^{-1/2}\frac{d\Delta(u)}{du}\right] = 
                                     2\pi\frac{d\phi(u)}{du}, 
\end{equation}
with solution 
\begin{equation}
\Delta(x^2) = 4\pi\int x^2 \phi(x^2) dx + c_1 x^3 + c_2.   \label{soln} 
\end{equation}
So one can in principle extract from Eq.~(\ref{soln}) 
a nonlocal gauge in which the bare propagator solves massless QED3.  
In practice, of course, determining the bare propagator gauge will not 
be a simple matter.  

Now consider the quenched theory, where the only free parameter is the 
coupling $e^2$ which for QED3 has dimensions of momentum.  
On dimensional grounds the chirally symmetric, Landau gauge solution 
to the fermion DSE must be of the form 
\begin{equation}
S_{\rm Landau}(x) = 
  \frac{\gamma\cdot x}{4\pi\left|x\right|^3} G(\left|x\right|e^2), 
\end{equation}
for some function $G$.  In any other covariant gauge, the propagator is 
given by the Euclidean LK transformation 
\begin{equation} 
S_\Delta(x) = S_{\rm Landau}(x)e^{e^2[\Delta(0) - \Delta(x^2)]}.  
\end{equation}
Specifically, if $\Delta_0$ is the gauge in which the DSE admits a bare 
propagator solution, then 
\begin{equation}
G(\left|x\right|e^2)\,e^{e^2[\Delta_0(0) - \Delta_0(x^2)]} = 1.  
\end{equation}
This can only be achieved if 
\begin{equation}
\Delta_0(0) - \Delta_0(x^2) = -\frac{\xi_0}{8\pi}\left|x\right|, 
\end{equation}
for some dimensionless constant $\xi_0$.  It follows\cite{BR93} that the gauge in which 
the propagator reduces to the the bare propagator is one of the usual 
covariant gauges 
\begin{equation}
\Delta_0(k^2) = -\frac{\xi_0}{k^4}, 
\end{equation}
and that in the general covariant gauge of Eq.(\ref{covgauge}), the fermion 
propagator is 
\begin{equation}
S(x) = \frac{\gamma\cdot x}{4\pi\left|x\right|^3} 
           e^{-e^2(\xi - \xi_0)\left|x\right|/8\pi}
\end{equation}
in coordinate space, or 
\begin{equation}
S(p) =  \frac{1}{i \gamma \cdot p} \left\{1 - \frac{e^2 (\xi - \xi_0)}
     {8 \pi p} \arctan \left[ \frac{8 \pi p}{e^2 (\xi - \xi_0)} \right] 
                   \right\},          \label{fermion3D}
\end{equation}
in momentum space.  

An analogous result is obtained for QED4 in ref.\cite{BKP98}, in which 
the perturbative correction to the vertex is used to model the unknown 
transverse contribution to the vertex function.  In our case we are faced 
with a similar problem: without knowing the transverse contribution 
$\bar{\Gamma}_\mu^{\rm T}$, we are unable to determine the constant $\xi_0$.  
The task of determining $\bar{\Gamma}_\mu^{\rm T}$ is a formidable one, 
and we have nothing more to say about it in this paper.  Instead, in the 
next section we concentrate on constructing a practical Ansatz for the 
remaining part of the vertex in an arbitrary covariant gauge.  

\setcounter{equation}{0}
\section{An Ansatz for part of the vertex}

In an arbitrary covariant gauge $\xi$, the vertex function can be 
decomposed according to 
\begin{equation}
\Gamma_\mu(p,q) = \hat{\Gamma}_\mu(p,q) + \bar{\Gamma}_\mu^{\rm T}(p,q), 
\end{equation}
where $\hat{\Gamma}_\mu(p,q)$ is obtained by LK transforming the bare 
vertex $\gamma_\mu$ from the special gauge $\xi_0$ obtained in the last 
section.  Likewise, $\bar{\Gamma}_\mu^{\rm T}(p,q)$ is obtained by LK 
transforming its value defined in Eq.~(\ref{barver}).  By making use of 
the LK transform of the vertex, namely
\begin{equation}
\Lambda_\mu(x,y,z;\Delta) = \Lambda_\mu(x,y,z;0) 
        e^{e^2\left[\Delta(0) - \Delta(x - y)\right]},  \label{LKver}
\end{equation}
where $\Lambda_\mu(p,q) = S(p)\Gamma_\mu(p,q)S(q)$, one readily checks that 
$\bar{\Gamma}_\mu^{\rm T}$ remains transverse and that $\hat{\Gamma}_\mu$ 
satisfies the WTI.  

The function $\hat{\Gamma}_\mu$ satisfies conditions (a) to (d) listed 
in the previous section.  In ref.\cite{BR93} these 
properties were exploited to place one further restriction on the form of 
the vertex Ansatz, which we shall henceforth refer to as the 
{\it transverse condition}: \\
(e) In the limit of zero fermion flavours and zero bare fermion mass, the 
part $\hat{\Gamma}_\mu$ of the fermion photon vertex Ansatz 
must be functionally dependent on the chirally symmetric fermion propagator 
$S(p) = 1/\left(i\gamma\cdot p A(p^2)\right)$ in such a way that
\begin{equation}
\int \frac{d^3k}{(2 \pi)^3} D_{\mu \nu}^{\rm T} (k - \ell) \gamma_\mu S(k) 
\hat{\Gamma}_\nu (k,\ell) = 0,  \label{tversecond}
\end{equation}
where $D_{\mu \nu}^{\rm T} (k)$ is defined by Eq.~(\ref{DTdef}).  

Although refs.\cite{BR93} and\cite{DMR94} only claim the transverse condition 
for the case $\xi_0 = 0$, we see by the following argument that it 
also applies to the current case.  In the gauge $\xi_0$, the transverse 
condition reduces to the identity Eq.~(\ref{angid}).  By Fourier 
transforming Eq.~(\ref{tversecond}) to the form 
\begin{equation}
\int d^3z\,D_{\mu \nu}^{\rm T}(z) \gamma_\mu \Lambda(x,y,z) = 0, 
\end{equation}
it is clear from Eq.~(\ref{LKver}) that the transverse condition is 
invariant with respect to LK transformations, and so should be true in 
any gauge.  

Following Eq.~(\ref{general}), we can write 
\begin{equation}
\hat{\Gamma}_\mu (p,q) = \Gamma_\mu^{\rm BC} (p,q) 
      + \hat{\Gamma}_\mu^{\rm T} (p,q).    \label{hatgeneral}
\end{equation}
By substituting Eq.~(\ref{hatgeneral}) into Eq.~(\ref{tversecond}) 
and noting that $\Gamma_\mu^{\rm BC}$ only has terms containing 
single $\gamma$ matrices we obtain the transverse condition in the form 
\begin{equation} 
\int \frac{d^3k}{(2 \pi)^3} \left\{D_{\mu \nu}^{\rm T}(k - \ell) 
  {\rm tr}\left[\gamma_\mu S(k) \Gamma_\nu^{\rm BC}(k,\ell) S(\ell)\right]
 + \frac{1}{(k - \ell)^2} F(k,\ell) \right\}= 0,   \label{tversecondition3}
\end{equation}
where 
\begin{equation}
F(k,\ell) =  {\rm tr}
  \left[\gamma_\mu S(k) \hat{\Gamma}_\mu^{\rm T}(k,\ell) S(\ell)\right].  
                           \label{Fdef}
\end{equation}

We now turn our attention to the photon polarisation scalar defined in 
Eq~(\ref{vacuum1}).  $\Pi(p^2)$ is a gauge invariant quantity\cite{LK56} 
and must remain so  in the limit $N_f\rightarrow 0$, that is, if calculated 
using a fermion propagator and vertex function obtained from the quenched 
version of the theory.  Furthermore, if we restrict our attention to that 
part $\hat{\Pi}(p^2)$ of the photon polarisation scalar arising from 
$\hat{\Gamma}_\mu$, we see from condition (d) that in 
the gauge $\xi_0$ it is given by the one-loop 
result\cite{ABKW86}
\begin{equation}
\Pi^{\rm one-loop}(p^2) = \frac{e^2}{8p}.    \label{vacuumLandau}
\end{equation}
We are thus led to one more constraint on the form of the Ansatz for 
$\hat{\Gamma}_\mu$, namely that: \\
(f) The part $\hat{\Pi}(p^2)$ of the photon polarisation 
scalar Eq.~(\ref{vacuum1}), calculated using $S(p)$ given by 
Eq.~(\ref{fermion3D}) and $\Gamma_\mu$ replaced by $\hat{\Gamma}_\mu$, 
must be equal to the one-loop result Eq.~(\ref{vacuumLandau}) 
for {\em all} choices of the gauge fixing parameter $\xi$.  

Inserting Eq.~(\ref{hatgeneral}) into the polarisation scalar 
Eq.~(\ref{vacuum1}) and using the WTI we obtain 
\begin{eqnarray}
\hat{\Pi} (p^2) & = & - \frac{e^2}{2 p^2} \int \frac{d^3q}{(2 \pi)^3} 
    \left\{{\rm tr} 
 \left[ \gamma_\mu S(q + \mbox{$\frac{1}{2}$}p) 
 \Gamma_{\mu}^{\rm BC} (q + \mbox{$\frac{1}{2}$}p,q - \mbox{$\frac{1}{2}$}p) 
   S(q - \mbox{$\frac{1}{2}$}p) \right] \right. \nonumber\\
&   & \left. \frac{3i}{p^2} {\rm tr} \left[ \gamma\cdot p \left( 
   S(q - \mbox{$\frac{1}{2}$}p) - S(q + \mbox{$\frac{1}{2}$}p)\right) \right] 
         + F(q + \mbox{$\frac{1}{2}$}p,q - \mbox{$\frac{1}{2}$}p) \right\}.    
                                     \label{vacuumscalar2}
\end{eqnarray}
It is clear that $\hat{\Gamma}_\mu^{\rm T}$ enters only through the 
function $F(k,\ell)$ defined by Eq.~(\ref{Fdef}).  
Furthermore, given any two vertex Ans\"atze satisfying conditions 
(a) to (e), and corresponding functions $F_1$ and $F_2$, we must have 
\begin{equation}
\int\frac{d^3k}{(2\pi)^3}\,\frac{1}{(k - \ell)^2} \left[
    F_1(k,\ell) - F_2(k,\ell) \right] = 0.  
\end{equation}
This follows from Eq.~(\ref{tversecondition3}), the first term in the 
integrand of which is a uniquely determined functional of $A(p^2)$.  
Integrating over $\ell$, making the change of 
variables $k_\mu = q_\mu + \frac{1}{2}p_\mu$, 
$\ell_\mu = q_\mu - \frac{1}{2}p_\mu$, and using 
Eq.~(\ref{vacuumscalar2}) then formally gives 
\begin{equation}
\int \frac{d^3p}{(2\pi)^3}\,\left[\hat{\Pi}_1(p^2) - 
                                   \hat{\Pi}_2(p^2)\right] = 0. 
\end{equation} 
We now take Ansatz number 1 to be any Ansatz satisfying only conditions (a) 
to (e), and Ansatz number 2 to be the `correct' Ansatz satisfying 
conditions (a) to (f).  We know in principle that Ansatz 2 exists.  It 
is simply the LK transform 
of the bare vertex in the gauge $\xi_0$ to any other gauge $\xi$.  
\footnote{An expression for the correct $\hat{\Gamma}_\mu$ is given 
in the Appendix to ref.\cite{BR93} in a form which is unfortunately of 
no immediate practical use.}  Since the transverse condition and the 
loop integral for the polarisation scalar are form invariant under 
the LK transformations, Ansatz 2 will then have properties (a) to (f).  
Ansatz number 1 must then entail a polarisation scalar satisfying  
\begin{equation}
\int_0^\infty dp \, p^2 \left[\hat{\Pi}(p^2) - \frac{e^2}{8p}\right] = 0, 
                              \label{psqpiint}
\end{equation}
provided this integral converges.  We note that the derivation of 
Eq.~(\ref{psqpiint}) requires a change of integration variable which, as 
we shall see, may or may not be valid, depending on the convergence 
properties of the integrals involved.  

In constructing an Ansatz for the vertex satisfying conditions (a) to (e), 
care must be taken to remember that not every function $F$ satisfying the 
transverse 
condition (e) corresponds to a viable vertex Ansatz satisfying criteria 
(a) to (d).  For instance, while one may be tempted simply to set the 
integrand of Eq.~(\ref{tversecondition3}) equal to zero, a straightforward 
substitution shows that this will not do, as the constraint $F(k,k) = 0$
following from Eq.~(\ref{tconstraint}) will then not be satisfied.  
One is led to take into account the general form of $\hat{\Gamma}^{\rm T}$ 
given by Eq.~(\ref{transverse}).  This gives 
\begin{eqnarray}
F(k,\ell) & = & \frac{4}{k^2 \ell^2 A(k^2)A(\ell^2)} 
  \left\{ \left[k^2\ell^2 - (k\cdot\ell)^2\right]
              \tilde{g}(k^2,\ell^2,k\cdot\ell) \right. \nonumber \\
& &  -2 \left[k^2\ell^2 + (k\cdot\ell)^2 - 
  (k^2 + \ell^2)k\cdot\ell\right]  g_3(k^2,\ell^2,k\cdot\ell)  \nonumber \\
& & \left.+ 2 (k^2 - \ell^2)k\cdot\ell  g_6(k^2,\ell^2,k\cdot\ell)  \right\},  
\end{eqnarray}
where 
\begin{equation}
\tilde{g}(k^2,\ell^2,k\cdot\ell) = g_8(k^2,\ell^2,k\cdot\ell) + 
                    (k^2 + \ell^2) g_2(k^2,\ell^2,k\cdot\ell).  
\end{equation} 

Dong et al.\cite{DMR94} proceed to satisfy the transverse condition by 
making the simplifying assumption that the $g_i$ be independent of 
$k\cdot\ell$.  The angular integral in Eq.~(\ref{tversecondition3}) can 
then be done analytically, giving the transverse condition as 
\begin{eqnarray}
0 & = & \int_0^\infty dk \frac{A(k^2) - A(\ell^2)}{k^2 - \ell^2} 
        \frac{1}{A(k^2)}    \times \nonumber \\
  & &  \left[I(k,\ell) - 
        2\frac{k^2 - \ell^2}{k^2 + \ell^2}I(k,\ell) f_6(k^2,\ell^2) 
         + J(k,\ell) \left(\mbox{$\frac{1}{2}$} \tilde{f}(k^2,\ell^2) 
                - f_3(k^2,\ell^2) \right) \right], \label{tversecondition4}
\end{eqnarray}
where $k$ and $\ell$ now mean $(k_\mu k_\mu)^{\frac{1}{2}}$ and 
$(\ell_\mu \ell_\mu)^{\frac{1}{2}}$ respectively, 
\begin{equation}
I(k,\ell) = \frac{(k^2 + \ell^2)^2}{16k\ell} \ln\left[\left(
  \frac{k + \ell}{k - \ell}\right)^2\right] - \frac{1}{4}(k^2 + \ell^2), 
\end{equation}
\begin{equation}
J(k,\ell) = \frac{(k^2 - \ell^2)^2}{16k\ell} \ln\left[\left(
  \frac{k + \ell}{k - \ell}\right)^2\right] - \frac{1}{4}(k^2 + \ell^2), 
\end{equation}
and we have set 
\begin{equation}
g_i(k^2,\ell^2) = \frac{A(k^2) - A(\ell^2)}{k^2 - \ell^2} f_i(k^2,\ell^2).   
\end{equation}
The $f_i$ are well defined if condition (d) holds.  

An Ansatz which satisfies conditions (a) to (e) can then be found by 
choosing functions $\tilde{f}$, $f_3$ and $f_6$ satisfying 
Eq.~(\ref{tversecondition4}) in such a way that they have no $\xi$ 
dependence other than a possible implicit dependence through $A$.  
The simplest way to achieve this, and the way chosen in ref.\cite{DMR94}, 
is to set the terms in square brackets in Eq.~(\ref{tversecondition4}) 
to zero.  Here we shall consider the one parameter family of Ans\"{a}tze 
\footnote{The choice $\tilde{f} = f_3 = 0$, $f_6 = 
\frac{1}{2}(k^2 + \ell^2)/(k^2 - \ell^2)$, namely the chiral limit of 
the Curtis-Pennington (CP) vertex\cite{CP91} was proposed in ref.\cite{BR93}.  
The chiral limit of the CP vertex is unacceptable as it violates condition  
(b) (the Ward identity)\cite{DMR94} and leads to a divergent integral for 
the polarisation scalar Eq.~(\ref{vacuumscalar2}).}
\begin{equation}
\tilde{f}(k^2,\ell^2) = -2(1 + \beta)\frac{I(k,\ell)}{J(k,\ell)}, 
\hspace{5 mm}  f_3(k^2,\ell^2) = -\beta\frac{I(k,\ell)}{J(k,\ell)}, 
  \hspace{5 mm}f_6(k^2,\ell^2) = 0.       \label{fansatz}
\end{equation}
The choice $\beta = 1$ gives the Ansatz proposed by Dong et al.\cite{DMR94}.  
One possible defect of this Ansatz for any value $\beta$ is that, for 
$k^2 = \ell^2$, but $k_\mu \ne \ell_\mu$, $\hat\Gamma_\mu$ has a logarithmic 
singularity.  (For $k_\mu = \ell_\mu$, the structure of the transverse 
tensors $T^i_\mu$ is such that $\hat\Gamma_\mu^{\rm T}(k,k) = 0$, but for 
$k_\mu$ and $\ell_\mu$ not parallel, no such cancellation occurs.) However, 
it is extremely difficult to construct functions $f_i$ satisfying 
Eq.~(\ref{tversecondition4}) without this defect, so we shall persevere 
with this family of Ans\"atze.  

\setcounter{equation}{0}
\section{Numerical evaluation of $\hat\Pi$}

We have numerically calculated $\hat{\Pi}$ using the vertex Ansatz 
containing the transverse piece specified 
by Eq.~(\ref{fansatz}).  This amounts to evaluating the integral 
\begin{eqnarray}
\hat{\Pi}(p^2) & = & -\frac{e^2}{\pi^2 p} \int_0^\infty dx \int_0^1 du 
  \frac{x^2}{\kappa^2\lambda^2}\left\{ x\left(3u - \frac{1}{u}\right) 
 \left[ \frac{\kappa^2}{A(p^2\lambda^2)} - 
    \frac{\lambda^2}{A(p^2\kappa^2)}  \right]    \right. \nonumber \\
 & & + xu \left[
    \frac{1}{A(p^2\kappa^2)} - \frac{1}{A(p^2\lambda^2)} \right]
     - \frac{1}{2}  \left[
    \frac{1}{A(p^2\kappa^2)} + \frac{1}{A(p^2\lambda^2)} \right]   
                          \label{vacuumscalar3} \\
 & & \left. + \frac{1}{\lambda^2 - \kappa^2}  \left[
    \frac{1}{A(p^2\kappa^2)} - \frac{1}{A(p^2\lambda^2)} \right]
          \left[x^2(1 - u^2)\tilde{f}(\kappa^2,\lambda^2) 
             + 2\left(x^2u^2 - \mbox{$\frac{1}{4}$}\right) 
      f_3(\kappa^2,\lambda^2) \right]\right\}, \nonumber
\end{eqnarray}
where 
\begin{equation}
\kappa^2  = x^2 + \mbox{$\frac{1}{4}$} + xu, \hspace{5 mm} 
\lambda^2 = x^2 + \mbox{$\frac{1}{4}$} - xu.  
\end{equation}
  
\begin{figure}
\epsfig{file=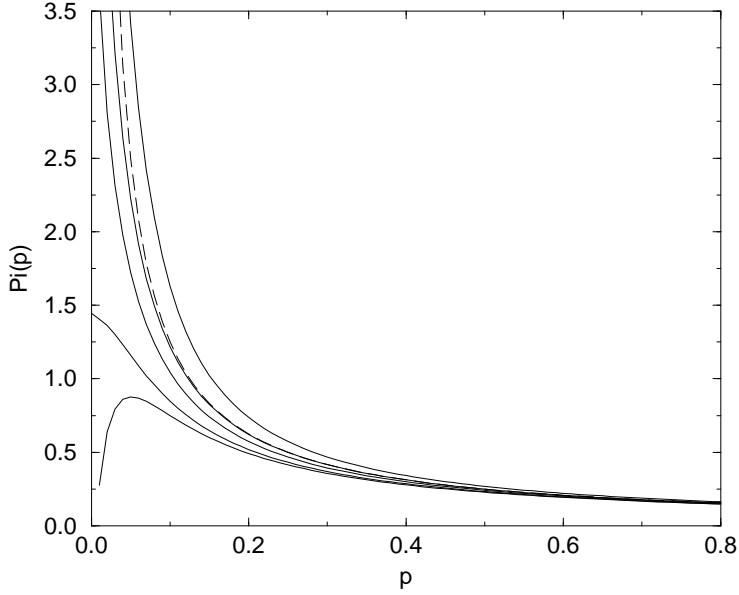,height=90mm}
\caption{That part $\hat{\Pi}(p^2)$ of the vacuum polarisation scalar arising 
from $\hat{\Gamma}_\mu^{\rm T}$.  The dashed curve is the  
one loop result Eq.~(\ref{vacuumLandau}), which is also the exact result 
in the gauge $\xi_0 = 0$. The solid curves are obtained using the 
Ansatz defined by Eq.(\ref{fansatz}) with (from bottom to top) 
$\beta$ = -1, -$\frac{2}{3}$, 0, 0.5942 and 2.  
\label{fig1}}
\end{figure}

The results are plotted in Fig.~\ref{fig1} for a range of values of 
$\beta$, together with the one-loop result Eq.~(\ref{vacuumLandau}).  If 
the fermion propagator of Eq.~(\ref{fermion3D}), which scales as 
\begin{equation}
S(p;e^2;\xi) = 
\frac{1}{e^2(\xi-\xi_0)} S\left(\frac{p}{e^2(\xi-\xi_0)};1;1 + \xi_0\right), 
\end{equation}
is used, the polarisation
scalar will scale as 
\begin{equation}
\hat{\Pi}(p;e^2;\xi) = \frac{1}{\xi-\xi_0}
    \hat{\Pi}\left(\frac{p}{e^2(\xi - \xi_0)};1;1 + \xi_0\right). 
\end{equation} 
In our numerical results we therefore only consider the case $e^2 = 1$, 
$\xi = 1 + \xi_0$.  

By expanding the integrand in Eq.~(\ref{vacuumscalar3}) before the factors 
of $p$ have been scaled out we obtain analytically the first couple 
of terms of the expansion of $\Pi(p)$ about $p = 0$: 
\begin{equation}
\hat\Pi(p) = 
\frac{1}{\xi - \xi_0}\left[1 - 2\left(\beta + \frac{2}{3}\right)
          \ln\left(\frac{4\pi p}{e^2(\xi - \xi_0)}\right) 
- \left(\beta + \frac{2}{9}\right) + O(p\ln p)  \right], 
\end{equation}
the first term being the contribution from the minimal Ball-Chiu part 
of the vertex.  This infrared behaviour is confirmed by the numerical 
results plotted in Fig.\ref{fig1}. 

On the other hand, inserting the expansion 
\begin{equation}
\frac{1}{A(q^2)} = 1 - \frac{e^2(\xi - \xi_0)}{16q} + 
    \left(\frac{e^2(\xi - \xi_0)}{8\pi q}\right)^2 - 
  \frac{1}{3} \left(\frac{e^2(\xi - \xi_o)}{8\pi q}\right)^4 + \ldots 
\end{equation} 
into Eq.~(\ref{vacuumscalar3}) we obtain an expansion for large $p$: 
\begin{eqnarray} 
\hat{\Pi}(p^2) & = & \frac{e^2}{8p} + 
\left(- \frac{\ln 2}{8\pi^2} + 6.768\times 10^{-3} 
   + 3.384\times 10^{-3} \beta \right) \frac{e^4(\xi - \xi_0)}{p^2} 
                                    + \nonumber \\
& &  \left(- \frac{1}{256\pi^2} + 3.957\times 10^{-4}\right) 
       \frac{e^6(\xi - \xi_0)^2}{p^3}  + O\left(\frac{1}{p^5}\right).  
\end{eqnarray}
The first term in each coefficient of this expansion arises from 
$\Gamma^{\rm BC}$ and has been calculated analytically, while 
the remaining terms arise from $\hat{\Gamma}^{\rm T}$ and have been 
calculated numerically.  

\begin{figure}
\epsfig{file=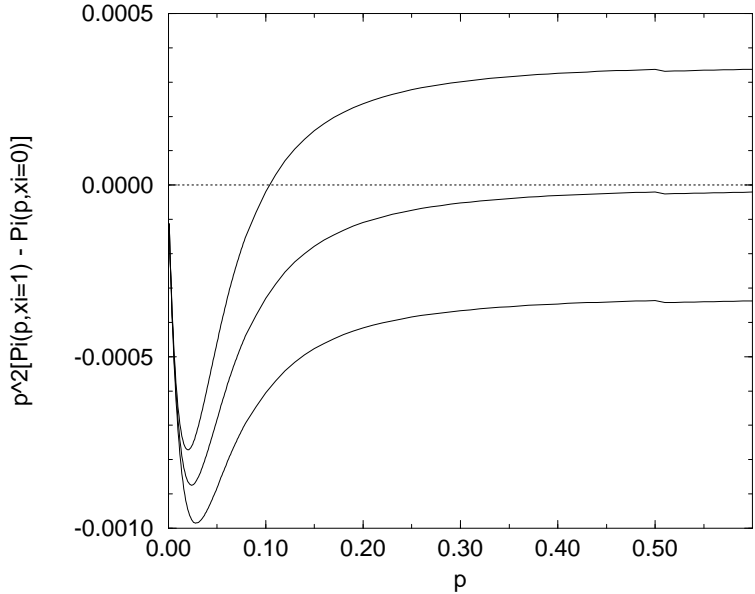,height=90mm}
\caption{A plot of the integrand $p^2[\hat{\Pi}(p^2) - e^2/8p]$ 
for $\beta = 0.5$ (bottom curve), 0.5942 (middle curve) and 0.7 
(top curve).  The middle value is that which gives a finite value 
for Eq.(\ref{psqpiint}).  
\label{fig2}}
\end{figure}

For the integral in Eq.~(\ref{psqpiint}) to be convergent, the 
coefficients of $1/p^2$ and $1/p^3$ should be zero.  The $1/p^2$ coefficient 
can be set to zero by choosing 
\begin{equation}
\beta = 0.5942.  \label{beta0}
\end{equation}
The coefficient of $1/p^3$ is zero to within the accuracy of our numerical 
calculations.  In Fig.\ref{fig2} we plot the integrand 
$p^2[\hat{\Pi}(p^2) - e^2/8p]$ 
for three values of $\beta$ including that given in Eq.~(\ref{beta0}).  
It is clear that the value of $\beta$ which renders the integral convergent 
leads to a nonzero value for the integral in Eq.~(\ref{psqpiint}).  We 
suspect that the cause of this discrepancy lies with the logarithmic 
singularity introduced into our Ansatz for $\hat\Gamma_\mu^{\rm T}$, 
which may have rendered the change of integration variables leading to 
Eq.~(\ref{psqpiint}) invalid.  Nevertheless, we see from Fig.\ref{fig1} 
that this choice of $\beta$ yields an excellent agreement with the exact 
one loop result over the range $p/e^2 > 0.1$.  

\setcounter{equation}{0}
\section{Summary and conclusions}

The problem of constructing a practicable Ansatz for the fermion-photon 
vertex function in QED3 has been considered.  We observe that the vertex 
contains a transverse part which cannot be determined by considering 
only the WTI and gauge covariance of single particle propagator DSEs.  
In order to progress further with determining this transverse part it 
will be necessary either to consider the vertex DSE or, following 
ref.\cite{BKP98}, to incorporate information from the perturbative loop 
expansion of the vertex.  These problems are not considered in this paper.  

In the process of carrying out our investigations, we have demonstrated 
the existence of a nonlocal gauge for which the chirally symmetric solution 
to the fermion DSE is the bare propagator.  In the case of the quenched 
theory $N_f\rightarrow 0$ this gauge is in fact one of the usual covariant 
gauges.  Unfortunately the actual gauge which renders the fermion propagator 
equal to the bare propagator gauge will remain unknown while the full 
transverse part of the vertex is undetermined.  

Nevertheless, we have explored the extent to which a vertex Ansatz can 
be constructed working with only single particle propagator DSEs in the 
quenched limit.  We write the vertex function as a sum of two pieces, 
one of which reduces to the bare vertex in the bare propagator gauge, 
and the other of which is a transverse piece containing information which can 
only be gleaned from higher $n$-point DSEs.  We give  a viable Ansatz for 
the first of these based on the work of Dong et al.\cite{DMR94}.  Our 
Ansatz is determined by the constraint that the known contribution to 
the photon polarisation scalar from this part of the vertex should be 
reproduced accurately.  Our Ansatz achieves this goal over practically 
the entire momentum range, failing only in the extreme infrared.  

\section*{Acknowledgement}

The authors gratefully acknowledge helpful discussions with P.\ Maris, 
M.\ R.\ Pennington, M.\ Reenders and A.\ H.\ Hams.  We are also grateful 
to the Special Research Centre for the Subatomic Structure of Matter, 
Adelaide, for hosting the Workshop on Nonperturbative Methods in Quantum 
Field Theory where part of this work was completed.  
\renewcommand{\theequation}{\Alph{section}.\arabic{equation}}
\setcounter{section}{1}
\setcounter{equation}{0}
\section*{Appendix}

Of the eight tensors spanning the the transverse part of the fermion-photon 
vertex function, only those consisting of terms containing an odd number of 
Dirac matrices are used in this paper.  They are given by 
\begin{eqnarray}
T_\mu^2(p,q) & = & \gamma\cdot(p + q)
   \left[p_\mu q\cdot(p - q)  - q_\mu p\cdot(p - q)\right]; \nonumber \\
T_\mu^3(p,q) & = & \gamma_\mu(p - q)^2 - (p - q)_\mu\gamma\cdot(p - q); 
                                                    \nonumber \\
T_\mu^6(p,q) & = & \gamma_\mu(p^2 - q^2) - (p + q)_\mu\gamma\cdot(p - q); 
                                                    \nonumber \\
T_\mu^8(p,q) & = & \mbox{$\frac{1}{2}$}[\gamma\cdot p \gamma\cdot q 
    \gamma_\mu - \gamma_\mu \gamma\cdot q \gamma\cdot p]  
\end{eqnarray}
Our definition of $T_\mu^2$ differs from that given in ref.\cite{DMR94}
by a minus sign.  We find this is necessary to reproduce Eqs.~(29) and 
(30) of ref.\cite{DMR94}.


\end{document}